\documentclass{pasj00}

\begin{document}
\SetRunningHead{F. Iwamuro et al.}{FMOS image-based reduction package}

\title{FIBRE-pac: FMOS image-based reduction package}

\author{
Fumihide \textsc{Iwamuro}\altaffilmark{1}
Yuuki \textsc{Moritani}\altaffilmark{1}
Kiyoto \textsc{Yabe}\altaffilmark{1}
Masanao \textsc{Sumiyoshi}\altaffilmark{1}
Kaori \textsc{Kawate}\altaffilmark{1}
Naoyuki \textsc{Tamura}\altaffilmark{2}
Masayuki \textsc{Akiyama}\altaffilmark{3}
Masahiko \textsc{Kimura}\altaffilmark{2}
Naruhisa \textsc{Takato}\altaffilmark{2}
Philip \textsc{Tait}\altaffilmark{2}
Kouji \textsc{Ohta}\altaffilmark{1}
Tomonori \textsc{Totani}\altaffilmark{1}
Yuji \textsc{Suzuki}\altaffilmark{1}
and
Motonari \textsc{Tonegawa}\altaffilmark{1}}
\altaffiltext{1}{Department of Astronomy, Kyoto University, Kitashirakawa, Kyoto, Japan}
\email{iwamuro@kusastro.kyoto-u.ac.jp}
\altaffiltext{2}{Subaru Telescope, National Astronomical Observatory of Japan, Hilo, HI, USA}
\altaffiltext{3}{Astronomical Institute, Tohoku University, Aramaki, Sendai, Japan}

\KeyWords{instrumentation: spectrographs --- methods: data analysis}

\maketitle

\begin{abstract}
The FIBRE-pac (FMOS image-based reduction package) is an IRAF-based reduction tool for the fiber multiple-object spectrograph (FMOS) of the Subaru telescope. To reduce FMOS images, a number of special techniques are necessary because each image contains about 200 separate spectra with airglow emission lines variable in spatial and time domains, and with complicated throughput patterns for the airglow masks. In spite of these features, almost all of the reduction processes except for a few steps are carried out automatically by scripts in text format making it easy to check the commands step by step. Wavelength- and flux-calibrated images together with their noise maps are obtained using this reduction package.

\end{abstract}
\section{Introduction}

FMOS \citep{kim10} is the near-infrared fiber multi-object spectrograph that has been in operation as one of the open-use instruments on the Subaru telescope since 2009. It can configure 400 fibers of $1\farcs 2$ aperture in a 30$'$ diameter field of view at the primary focus. The 400 infrared spectra in two groups are taken by two spectrographs called IRS1 and IRS2 (infrared spectrograph 1 and 2) in either of two modes: a low-resolution mode with a spectral resolution of $\Delta\lambda \sim$ 20 \AA\ in the 0.9-1.8 $\mu$m wavelength range, and a high-resolution mode of $\Delta\lambda \sim$ 5 \AA\ in one of the quarter wavelength ranges of 0.90-1.14, 1.01-1.25, 1.11-1.35, 1.40-1.62, 1.49-1.71, or 1.59-1.80 $\mu$m. The bright OH-airglow emission lines are masked by the mask mirror \citep{iwa01} installed in these spectrographs.

The individual tools for image reduction in this package were developed during engineering and guaranteed-time observations since 2008, in conjunction with other operation and reduction software. Although these tools were not developed as part of the public reduction software of FMOS, they are now stable, thus they have earned the name of ``FIBRE-pac'' (FMOS Image-Based REduction package). The basic concepts underlying the package are as follows:
\begin{enumerate}
\item Most of the reduction is processed by IRAF.
\item Several complicated steps are processed by original software written in C using the cfitsio library \citep{pen99}.
\item Almost all of the reduction processes are automated using script files in text format.
\item Modification of the text files is done by general UNIX commands.
\item The original 2-dimensional information is kept as far as possible throughout the reduction processes.
\end{enumerate}
These concepts enable easy implementation and open processing without the inconvenience of licensed or black box parts and ensure traceable operation with visual confirmation. The 2-dimensional information has advantages not only in filtering out unexpected noise using their small sizes but also on detection of faint emission-features. In this paper, we describe the reduction process of the FMOS images based on its Aug.9-2011 version, taking care of complex conditions in the infrared, using multiple fibers and the OH-suppressed spectrograph.

\section{The images}
The FMOS images are acquired by a uniform interval non-destructive readout technique called ``Up-the-Ramp sampling'' (hereafter ``ramp sampling'' for short). A typical exposure of 900 seconds consists of 54 images. After an exposure is finished, a final frame (treated as a ``raw image'' in the reduction process) is created where the count in each pixel is calculated by performing a least squares fit to the signal count of 54 images. In addition to suppressing readout noise, the advantages of the ramp sampling are 1) saturated pixels can be estimated from the counts prior to saturation, and 2) cosmic-ray events can be detected as an unexpected jump in the counts and removed from the final frame \citep{fix00}. The detection threshold of the cosmic-ray events has been currently set to 10$\sigma$ in an empirical way based on the real FMOS images. For IRS1, the fit and the cosmic-ray rejection are done during the ramp sampling, so that the final frame is ready as the exposure finishes. For IRS2, however, the nonlinear bias variation prevents fitting the slope during the exposure. Instead, the ramp fitting is executed after the simple background subtraction (cf. \S 3.2) has been performed for all of the images taken during the sampling. For example, 54 background-subtracted ramp images are prepared prior to fitting. Consequently the nonlinear bias component is subtracted together with the background photons.

\begin{figure}
  \begin{center}
    \FigureFile(80mm,80mm){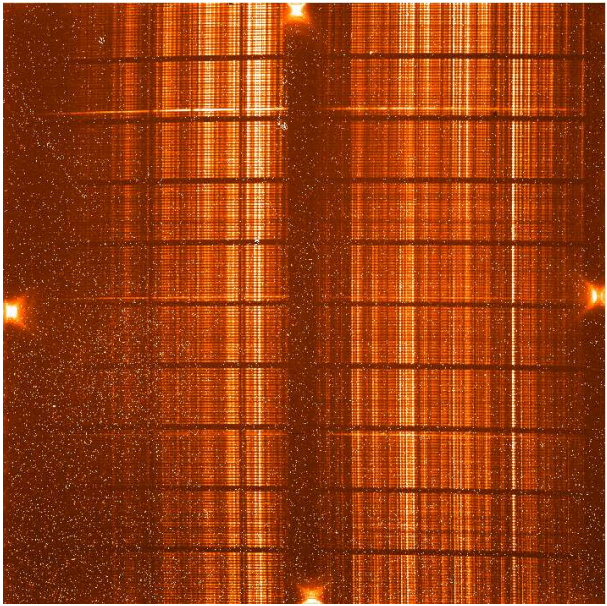}
  \end{center}
  \caption{Raw image taken in the IRS1 low-resolution mode with a ramp readout of 54 times using the HAWAII-2 2k$\times$2k detector (corresponding to a 900-s exposure). The left half of the image consists of spectra in the $J$ band, while the right half is in the $H$ band. The typical FWHM of the each fiber spectrum is 5 pixels with a pitch of 10 pixels between the spectra.
}\label{fig:raw_image}
\end{figure}

Figure \ref{fig:raw_image} shows the resulting ``raw image'' after ramp sampling 54 times using IRS1. Raw images for IRS2 do not exist for the reasons explained above. The raw images have the following components:
\begin{enumerate}
\item Thermal background ($\sim$300 e/900 s)
\item Suppressed OH-airglow ($\sim$600 e/900 s on average)
\item Remaining cosmic ray events ($\sim$200 e/event)
\item Read noise ($\sim$10 e rms/54 ramp readout)
\item Object ($\sim$30 e/900 s for a 20 mag(AB) object)
\item Dark current ($<$10 e/900 s)
\item Bias offset ($<$10 e/exposure)
\item Cross-talk between quadrants ($\sim$0.15\% of the count)
\item Bad pixels having no efficiency, unusual large dark current, or too large noise ($\sim$0.2\% of the pixels)
\item On-chip amplifier glow at the corners of the four quadrants ($\sim$2000 e/54 ramp readout)
\item Unknown external noise in a stripe pattern (quite rare and the amplitude is $\sim$100 e/900 s at most)
\item Point spread function (PSF) of the fiber of 5 pixels with a pitch of 10 pixels between spectra (having almost no cross talk of the spectra on either side)
\item Position of an individual fiber along the slit
\item Optical distortion
\item Quantum efficiency and its pixel-to-pixel variation (``flat'' pattern)
\item System throughput variation with wavelength
\item Atmospheric transmittance
\item Intrinsic absorption and emission of the objects
\end{enumerate}
The thermal background can be removed in the initial subtraction described in subsection 3.2. The 2nd strongest component, residual airglow, can be subtracted by interpolation of the background of other fibers after the optical distortion is corrected using a dome-flat image taken in the same observation mode. The remaining cosmic-ray and other strong noise features having a size smaller than the PSF of a fiber are removed in three subsequent stages with different threshold counts. Besides the science frames, the following images are necessary for this reduction process.
\begin{description}
\item[Detector-flat image]: The homogeneous thermal image of the black board attached to the entrance window of the camera dewar. This image gives the pixel-to-pixel quantum efficiency ratio of the detector. More than 30 sets of the ramp sampling frames are averaged to ensure the high signal-to-noise (S/N) ratio. This image is included in the reduction package together with the other standard images such as a bad pixel mask.
\item[Bad pixel mask]: The distribution of bad pixels in the detector selected through the reduction process of the detector-flat image.
\item[Dome-flat image]: The dome-flat spectra taken before or after observation. This image is used for measurements of the distortion parameters as well as the relative throughput correction of the scientific images.
\item[Th-Ar spectral image]: The emission-line spectra for wavelength calibration taken just before the exposures of dome-flat frames.
\end{description}
The details of the reduction processes using these images are described in the following section.

\section{Reduction Process}
\subsection{Preparatory processing}
Before performing a reduction of the science frames, a preparatory process is applied to the dome-flat and Th-Ar spectral images to determine the optical distortion and the wavelength calibration of the image. 

First, Flat fielding: The dome-flat image is divided by the detector-flat image to remove differences in the quantum efficiency between pixels. 

Second, Bad pixel correction: The registered and temporally prominent bad pixels, which are picked up by subtraction of 3$\times$3 median filtered image, are replaced with an interpolated value from the surrounding pixels. 

\begin{figure*}
  \begin{center}
    \FigureFile(80mm,79mm){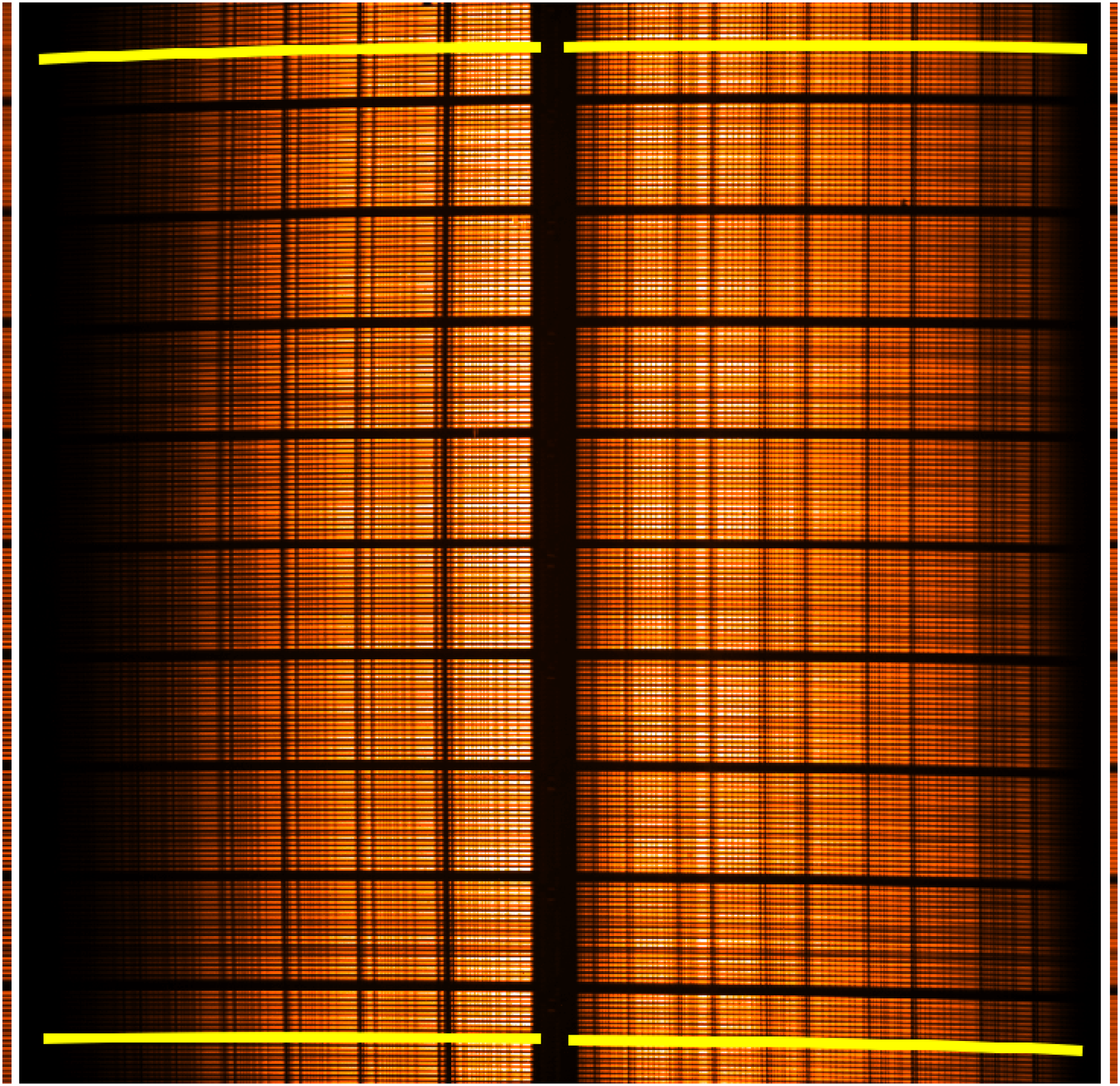}
    \FigureFile(80mm,80mm){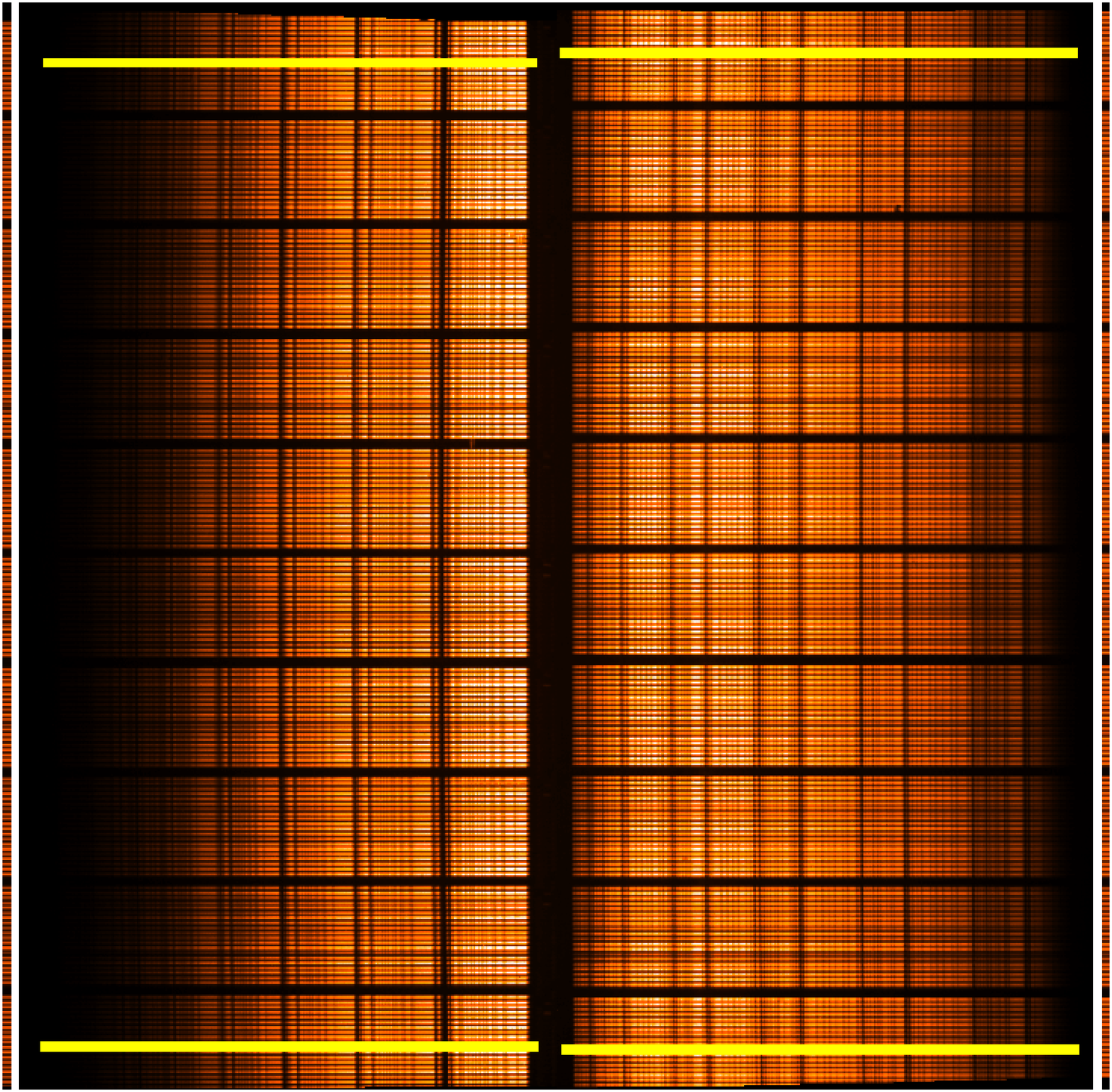}
  \end{center}
  \caption{Correction of the spatial distortion of the dome-flat image. The $y$-axis of the left-hand image is converted to straighten the spectra, as shown at right. The thick lines indicate the shape of the spectra at the top and bottom edges of the detector. The left and right half regions are separately corrected in this low-resolution mode, because the corresponding mask mirrors are separated \citep{kim10}. The projected images of each half region along the $x$-axis are shown along the sides of the image.
}\label{fig:spacial_distortion}
\end{figure*}

\begin{figure}
  \begin{center}
    \FigureFile(80mm,18mm){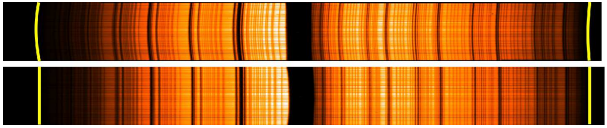}
  \end{center}
  \caption{Correction of the spectral distortion of the combined 1D image of the dome flat. The $x$-axis of the combined 1D image (at top) is converted using $x^\prime = x+(ax+b)$, in which $a$ and $b$ are coefficients of a fourth-order polynomial function in $y$. The images of the airglow masks are consequently straightened in the corrected image (at bottom). The thick lines indicate the shape of the airglow mask at the left and the right edges of the detector.}\label{fig:spectral_distortion}
\end{figure}

Third, Correction of the spatial distortion: The $y$-axis of the image is converted using $y^\prime = y+(a_1y+a_2)x^2+(b_1y+b_2)x$. (Here, the origin is at the center of the image.) The four parameters are chosen to make the PSF amplitude in the image projected along the $x$-axis maximum (cf. figure \ref{fig:spacial_distortion}). In this modification process, the dome-flat spectrum from each fiber with 9 pixels in width is converted into a parallel line to make a ``combined 1D image'' (as in figure \ref{fig:spectral_distortion}) that includes the pattern of the airglow mask. 

Fourth, Correction of the spectral distortion: The $x$-axis of the combined 1D image is converted using $x^\prime = x+(ax+b)$, in which the two parameters are determined for each line so as to minimize the shift and the magnification difference in the mask pattern. The 200 resulting $a$ and $b$ parameters are fitted with fourth-order polynomials in $y$ to remove the local matching error in the pattern. This conversion process makes the mask pattern straight in the column direction (cf. figure \ref{fig:spectral_distortion}), which is necessary for better subtraction of the residual airglow lines in the science frames. 

\begin{figure}
  \begin{center}
    \FigureFile(80mm,18mm){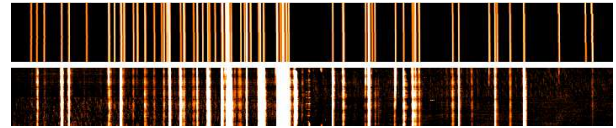}
  \end{center}
  \caption{Combined 1D image of the Th-Ar spectra (at bottom) compared with an artificial image based on the known wavelengths of the Th-Ar emission lines (at top). The agreement in the position of the lines shows that the conversion from the column numbers to wavelengths is correct.}\label{fig:thar_lamp}
\end{figure}

These parameters obtained from the dome-flat image are also applied to the Th-Ar spectral image to make a combined 1D image, as shown in figure \ref{fig:thar_lamp}. Although the slit image of each emission line and of the airglow masks should be parallel, alignment errors cause small differences between them.

Finally, Wavelength calibration: The correlation between the observed wavelength and the pixels of each spectrum are determined, as represented by $\lambda = px^3+qx^2+rx+s$. The resulting $p$, $q$, $r$, and $s$ coefficients are used to calculate the corresponding wavelengths in each spectrum without changing them, because the $x$ positions of the individual fibers in the slit may exhibit some scatter due to imperfect fiber alignment along the slit. The results of the wavelength calibration are confirmed by comparing the reduced Th-Ar spectra with an artificial image based on the known wavelengths of the Th-Ar emission lines (cf. figure \ref{fig:thar_lamp}). The typical calibration error of the spectra is less than 1 pixel, corresponding to 5 \AA\ or 1.2 \AA\ in the low- or high-resolution mode, respectively.

\subsection{Initial background subtraction}
\begin{figure*}
  \begin{center}
    \FigureFile(50mm,50mm){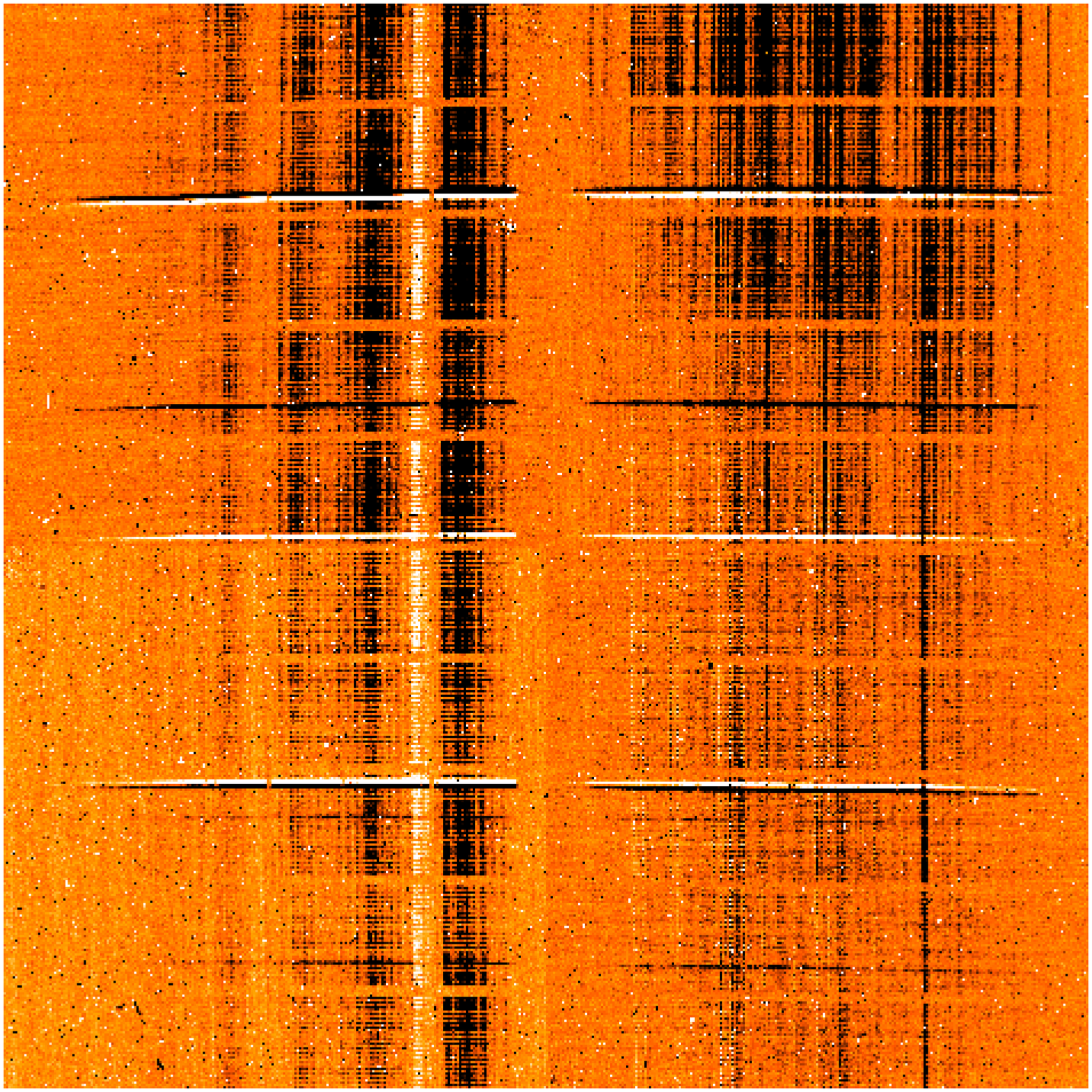}
    \FigureFile(50mm,50mm){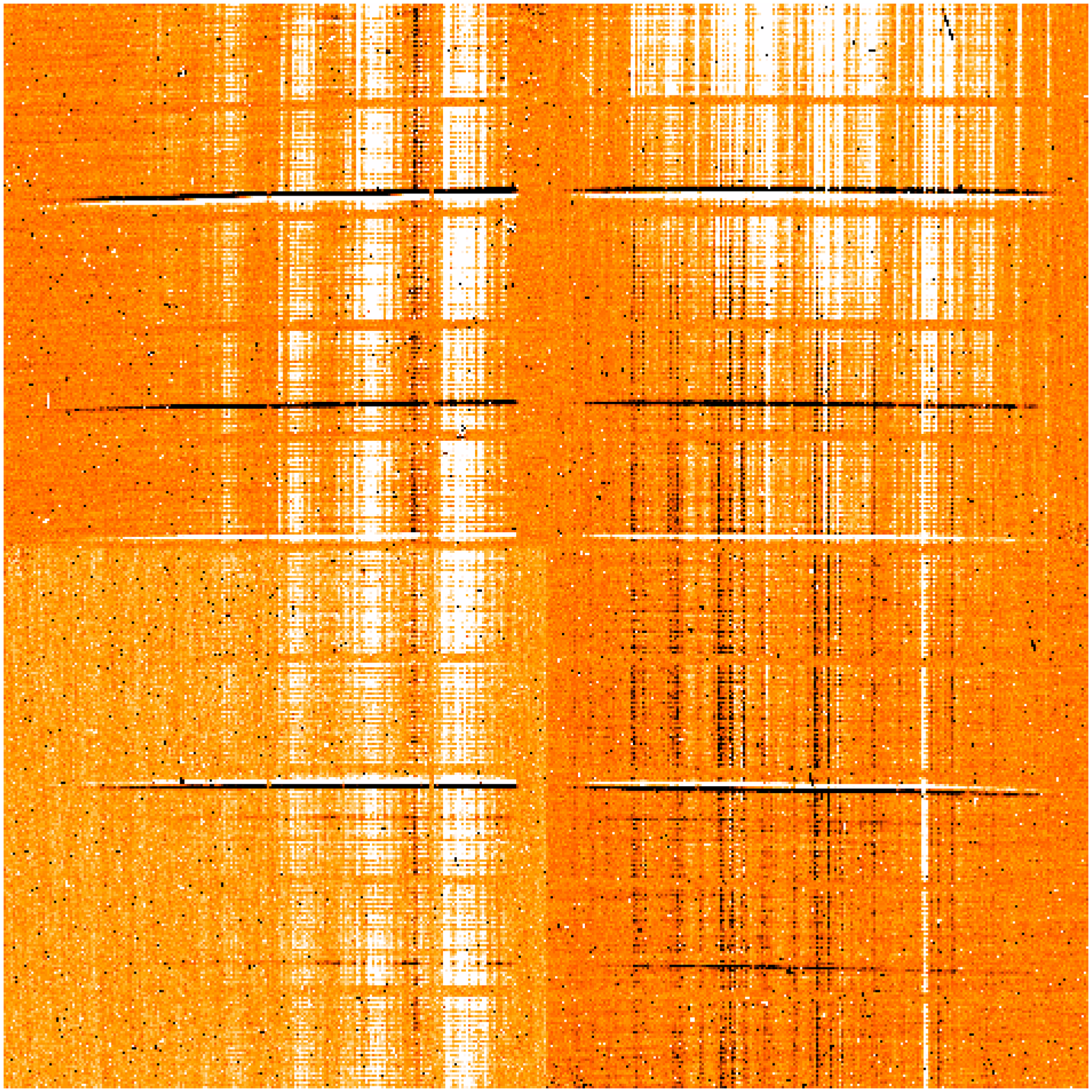}
    \FigureFile(50mm,50mm){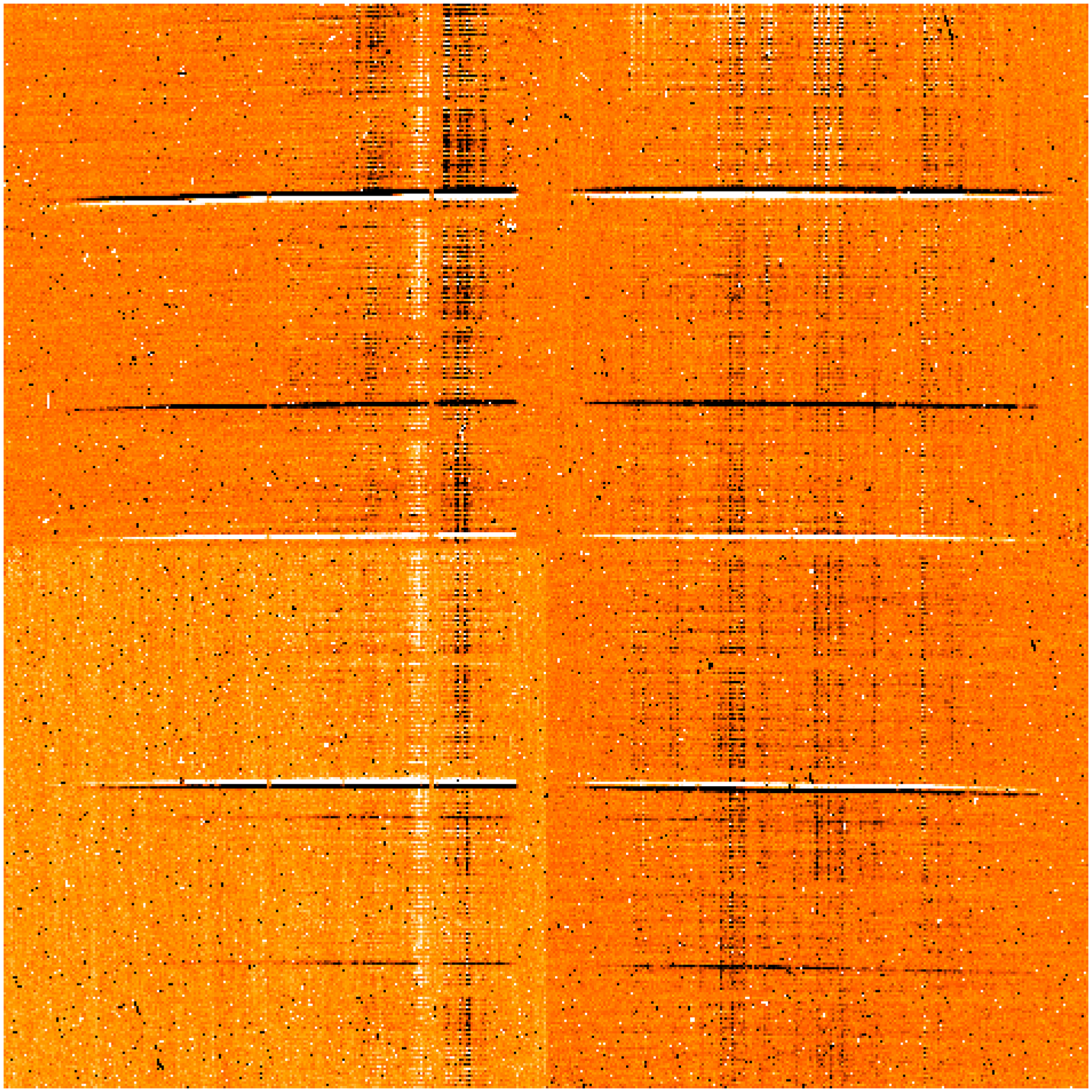}
  \end{center}
  \caption{Canceling the residual OH-airglow emission. Left panel: Sky-subtracted image using the previous image as the sky image ($A_n-B_{n-1}$). Center panel: Sky-subtracted image using the next image as the sky image ($A_n-B_n$). Right panel: Merged image ($A_n-B$) with a weight of $w=0.472$ in which the negative spectra cannot be used because they have merged information with various weights.}\label{fig:initial_subtraction}
\end{figure*}

Since the usual FMOS observations are carried out using an $ABAB$ nodding pattern of the telescope, $A-B$ simple sky subtractions can be performed using two different sky images: $A_n-B_{n-1}$ and $A_n-B_n$ (where $A_n$ denotes the image taken at position $A$ of the $n$-th pair). For IRS2, these sky-subtracted frames are calculated by the ramp fitting algorithm applied to the sky-subtracted sub-frames. When the brightness of the OH airglow varies monotonically, most of it can be canceled by merging these two images according to $A_n-B=(A_n-B_n)w+(A_n-B_{n-1})(1-w)$. The weight $w$ is chosen to make the sum of the absolute count $|A_n-B|$ a minimum within the range $-1<w<2$. Figure \ref{fig:initial_subtraction} shows a pair of sky-subtracted images and the merged one. Typically, the weight $w$ is equal to about 0.5. 

\subsection{Corrections of detector cross talk, bias difference, and bad pixels}
After the initial background subtraction, cross talk is removed by subtracting 0.15\% for IRS1 and 1\% for IRS2 from each quadrant. Next, the bias difference between the quadrants (as indicated in figure \ref{fig:initial_subtraction}, the lower left quadrant tends to show a higher bias level) is corrected to make the average over each quadrant equal. After flat fielding using the detector flat image, the registered and temporally prominent bad pixels are rejected, together with the four adjacent pixels by interpolating the surrounding pixels.

\subsection{Distortion correction and residual sky subtraction}
\begin{figure*}
  \begin{center}
    \FigureFile(50mm,50mm){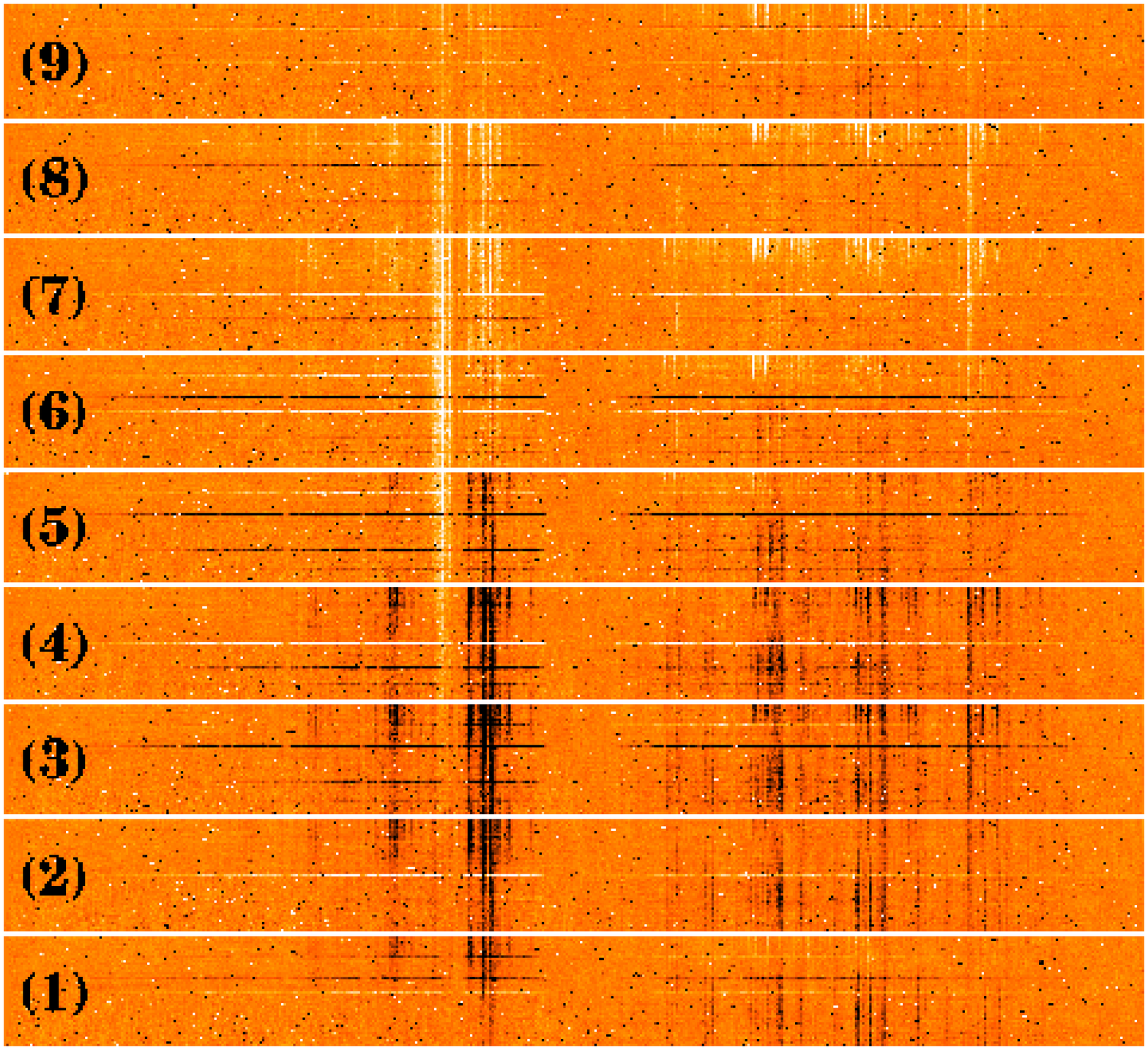}
    \FigureFile(50mm,50mm){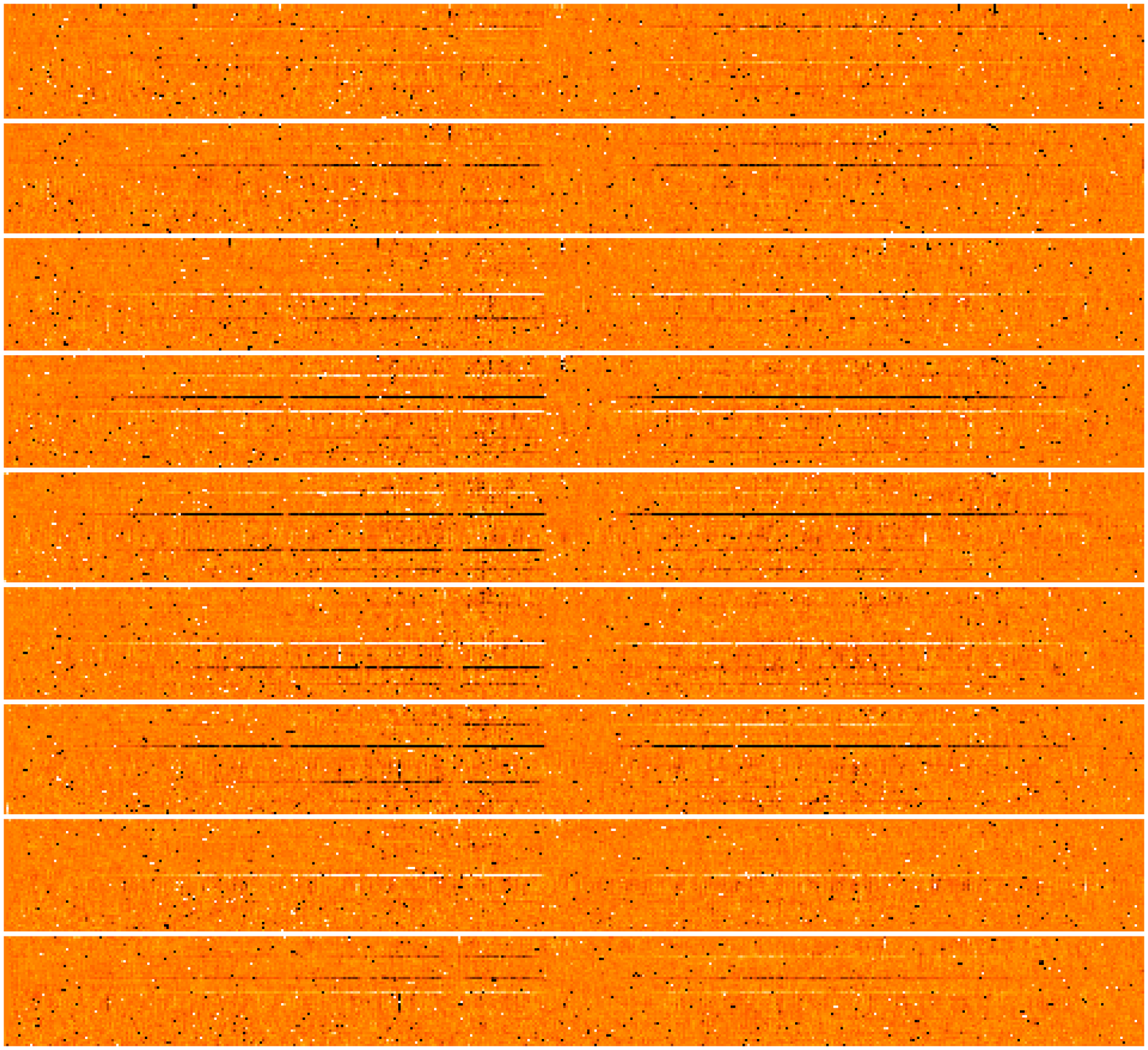}
    \FigureFile(50mm,50mm){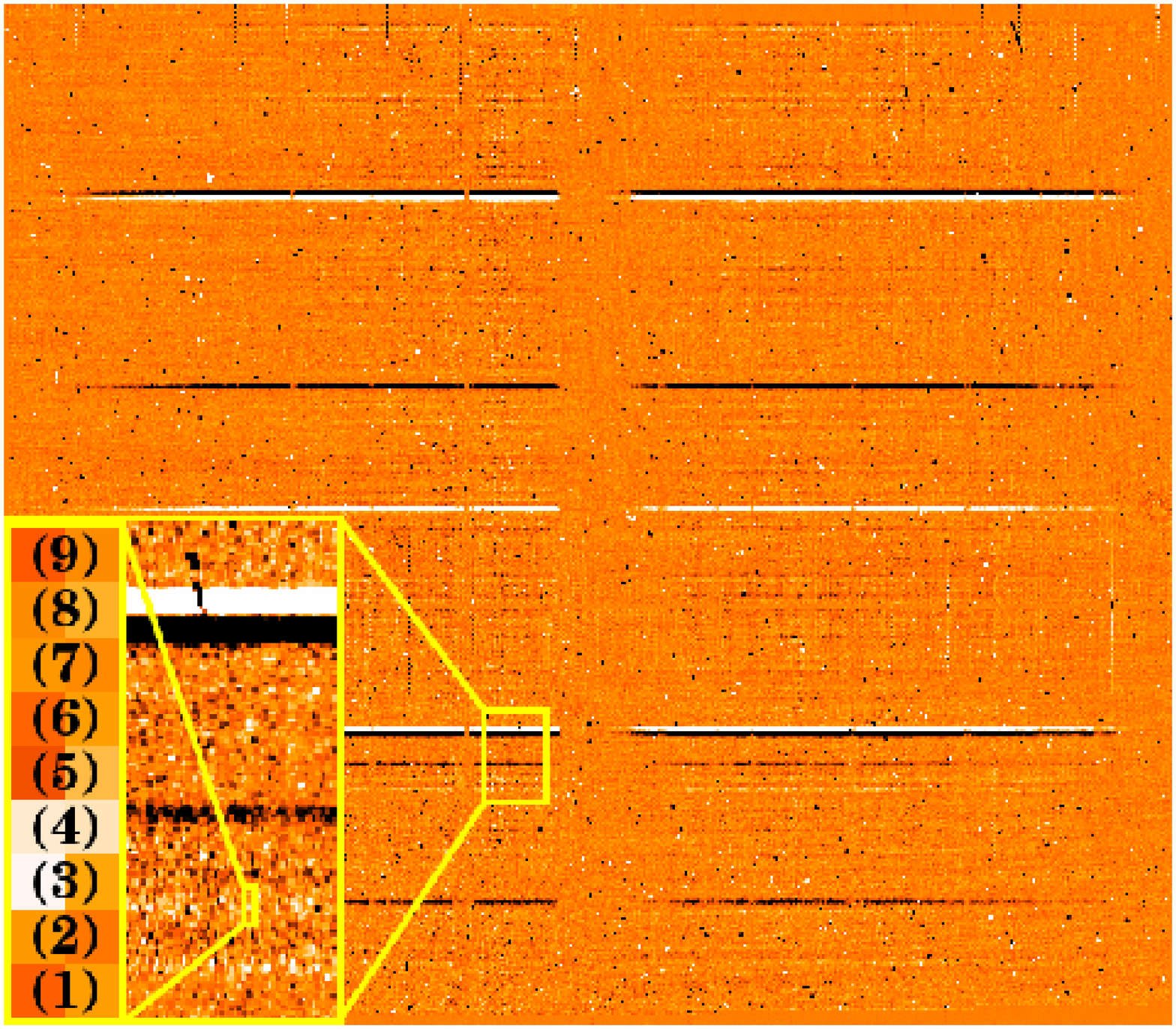}
  \end{center}
  \caption{Residual background subtraction and recomposition process. Left panel: A set of 9 combined 1D images from different part of the PSF. Center panel: Residual subtracted images by performing fits along the columns in each image. Right panel: Recomposed image from the 9 residual subtracted images.}\label{fig:residual_subtraction}
\end{figure*}

The processed image is converted into a combined 1D image based on the distortion parameters obtained in the preparatory reduction process. However, only one line of each spectrum (9 pixels in width) is extracted instead of summing the 9 lines. One can thereby make a set of 9 combined 1D images each of which consists of a different part of each spectrum (cf. figure \ref{fig:residual_subtraction}). In other words, the PSF of a fiber is divided into 9 pieces, only one of which is used in a combined 1D image. The flexure or temperature change in the spectrograph causes a small difference between the dome-flat and the scientific images along the vertical direction in position. This difference is corrected using the vertical position of the spectra of bright stars. In this way, the counts of the residual sky becomes smooth along the columns after the relative throughput correction of the fibers. The residual airglow lines are fitted and subtracted in these images, and then the relative throughput difference is multiplied to restore the noise level back to the original state. The residual subtracted images are recombined to form an image in which the PSF of the fiber is determined. Medium-level bad pixels having smaller size than the PSF of a fiber are replaced at the end of this process.

\subsection{Combine and residual background subtraction}
\begin{figure*}
  \begin{center}
    \FigureFile(50mm,50mm){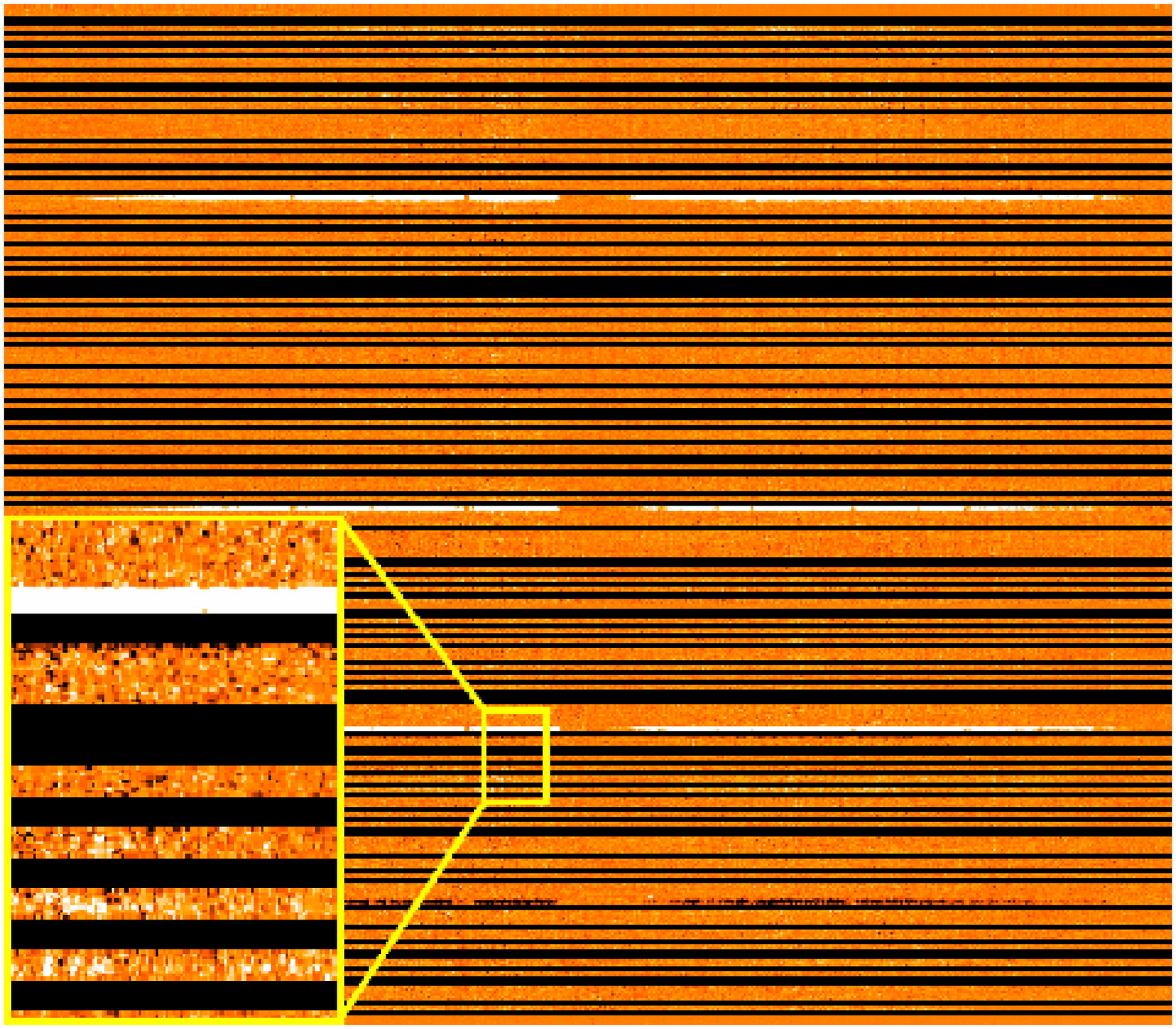}
    \FigureFile(50mm,50mm){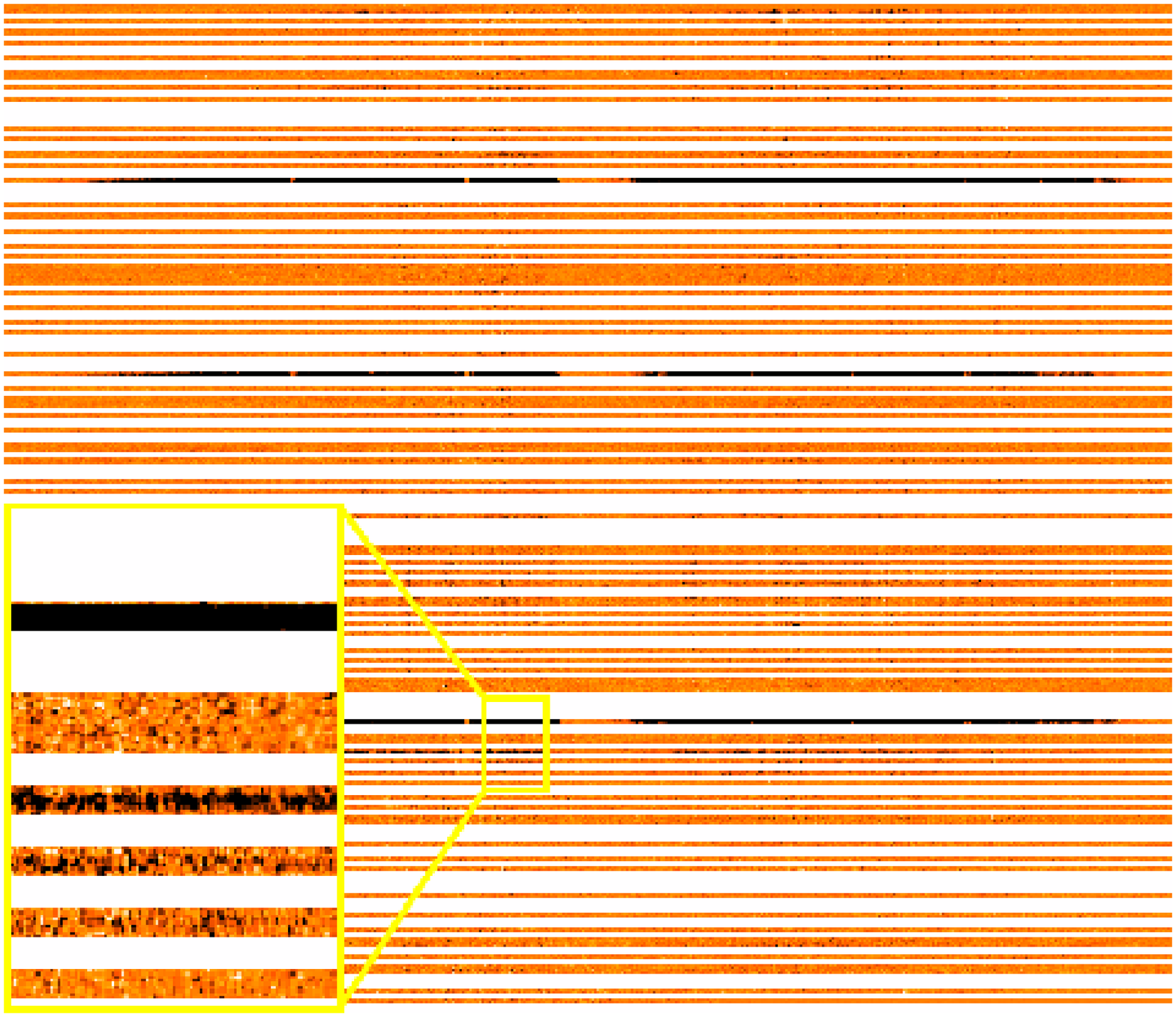}
    \FigureFile(50mm,50mm){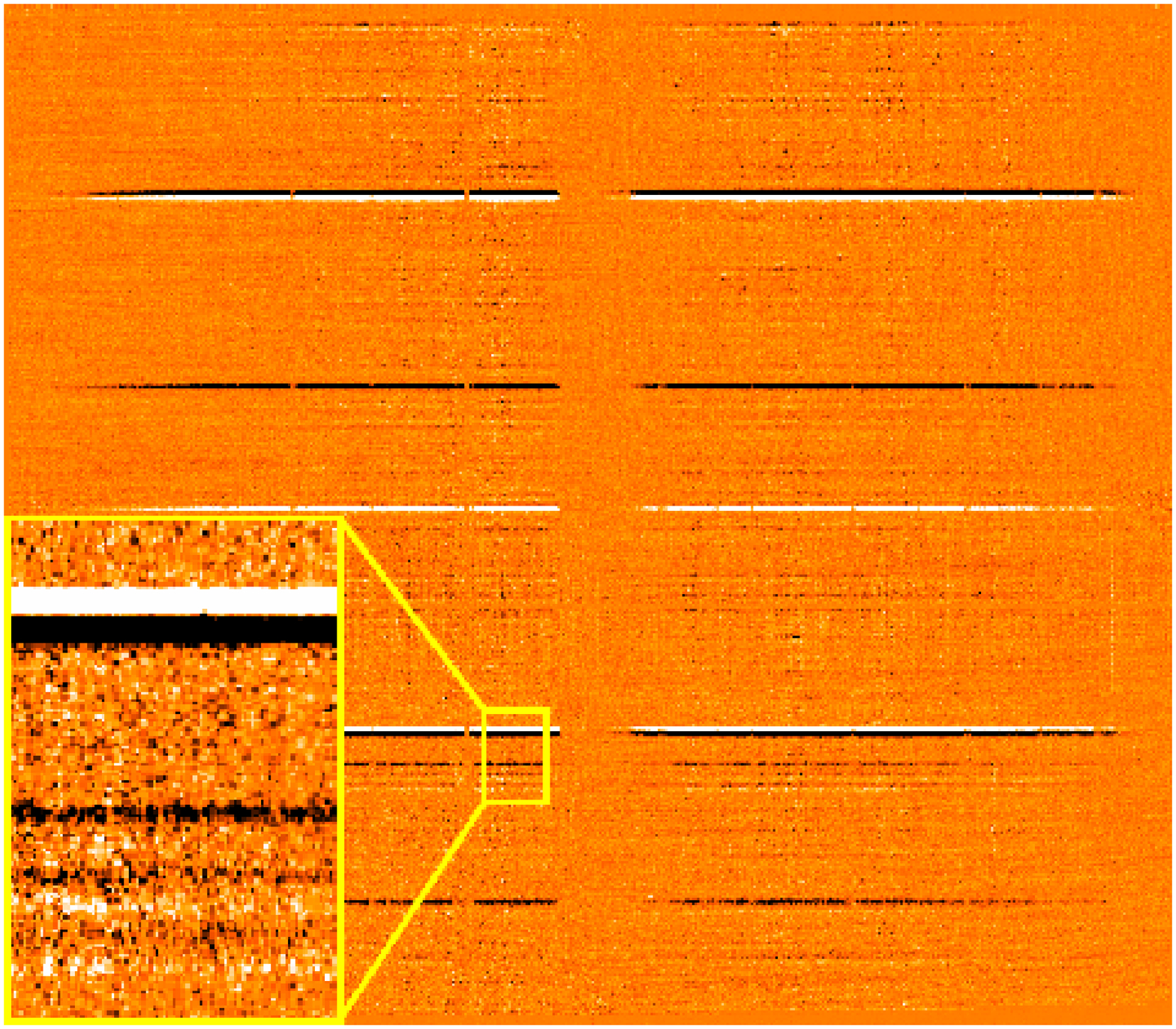}
  \end{center}
  \caption{Merging the CBS images from positions $A$ and $B$. Left panel: Positive spectra in the reduced $A_n-B$ image in which the removed negative (position $B$) spectra contain merged information (cf. figure \ref{fig:initial_subtraction}). Center panel: Negative spectra in the reduced $A-B_n$ image in which the removed positive (position $A$) spectra are merged one. Right panel: Merged $A_n-B_n$ image.}\label{fig:merge_pair}
\end{figure*}

There are two ways to observe scientific targets with the $ABAB$ nodding pattern: ``normal beam switching'' (NBS), in which all targets are observed at position $A$ (all the fibers are supposed to observe ``blank'' sky at position $B$), and ``cross beam switching'' (CBS), in which less than half of the fibers are allocated to the targets at position $A$ while the others are at position $B$. In the CBS observation mode, the same targets appear in both images collected in positions $A$ or $B$. However, the target spectra in position $B$ are merged to minimize the absolute flux of the residual airglow lines in the initial background subtraction process. Thus the same reduction process has to be applied to the images in position $B$, replacing the negative spectra of position $A$ with the corresponding parts of the position $B$ image (as shown in figure \ref{fig:merge_pair}). The merged images (or all of the position $A$ images in the NBS mode) are then combined into one averaged image.

Finally, the combined image is divided into a set of 9 combined 1D images again, in order to perform a fine subtraction of the residual background. The recomposed image also goes through a bad pixel rejection process.

\subsection{Mask edge correction and CBS combine}
\begin{figure}
  \begin{center}
    \FigureFile(80mm,18mm){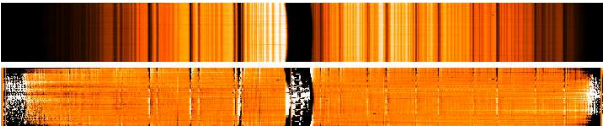}
  \end{center}
  \caption{Relative dome-flat image to correct the throughput differences between spectra. Top panel: Flux-normalized dome-flat spectra in which the throughput differences of the fiber have been removed. Bottom panel: Relative dome-flat image created by normalizing the top image along the columns. This image is displayed within a range of 0.8 (black) to 1.2 (white) to emphasize the relative differences.}\label{fig:relative_domeflat}
\end{figure}

\begin{figure*}
  \begin{center}
    \FigureFile(50mm,50mm){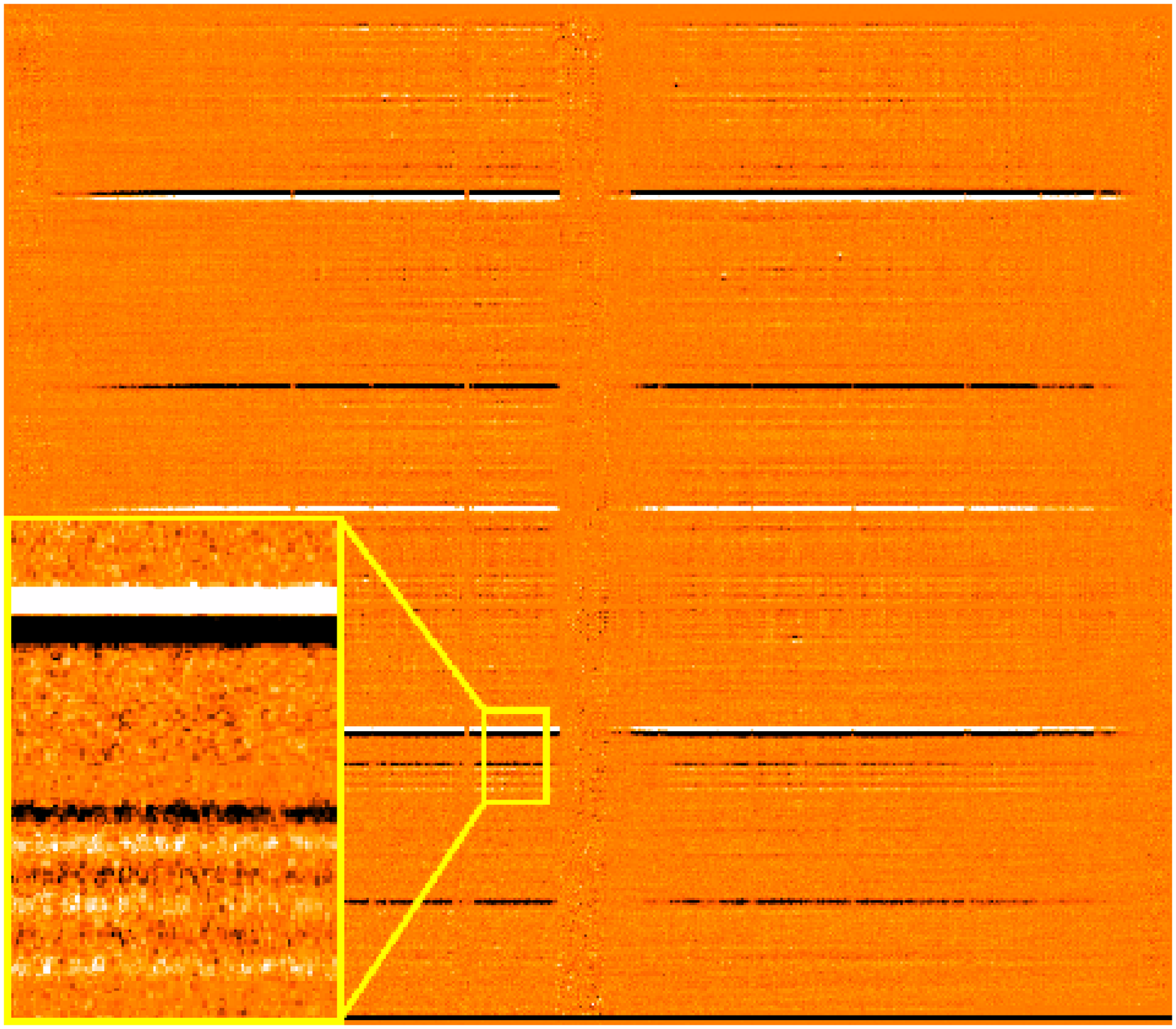}
    \FigureFile(50mm,50mm){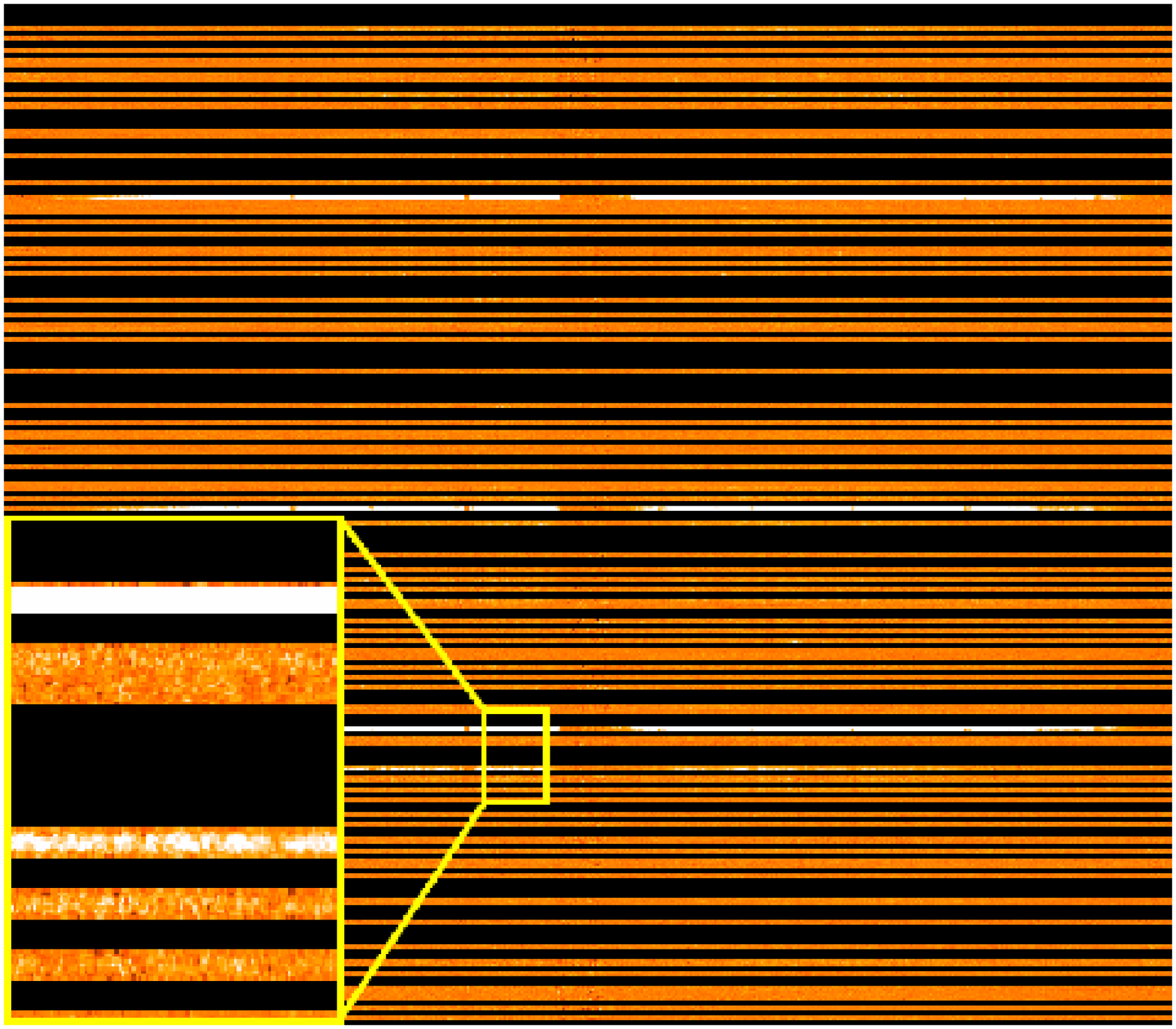}
    \FigureFile(50mm,50mm){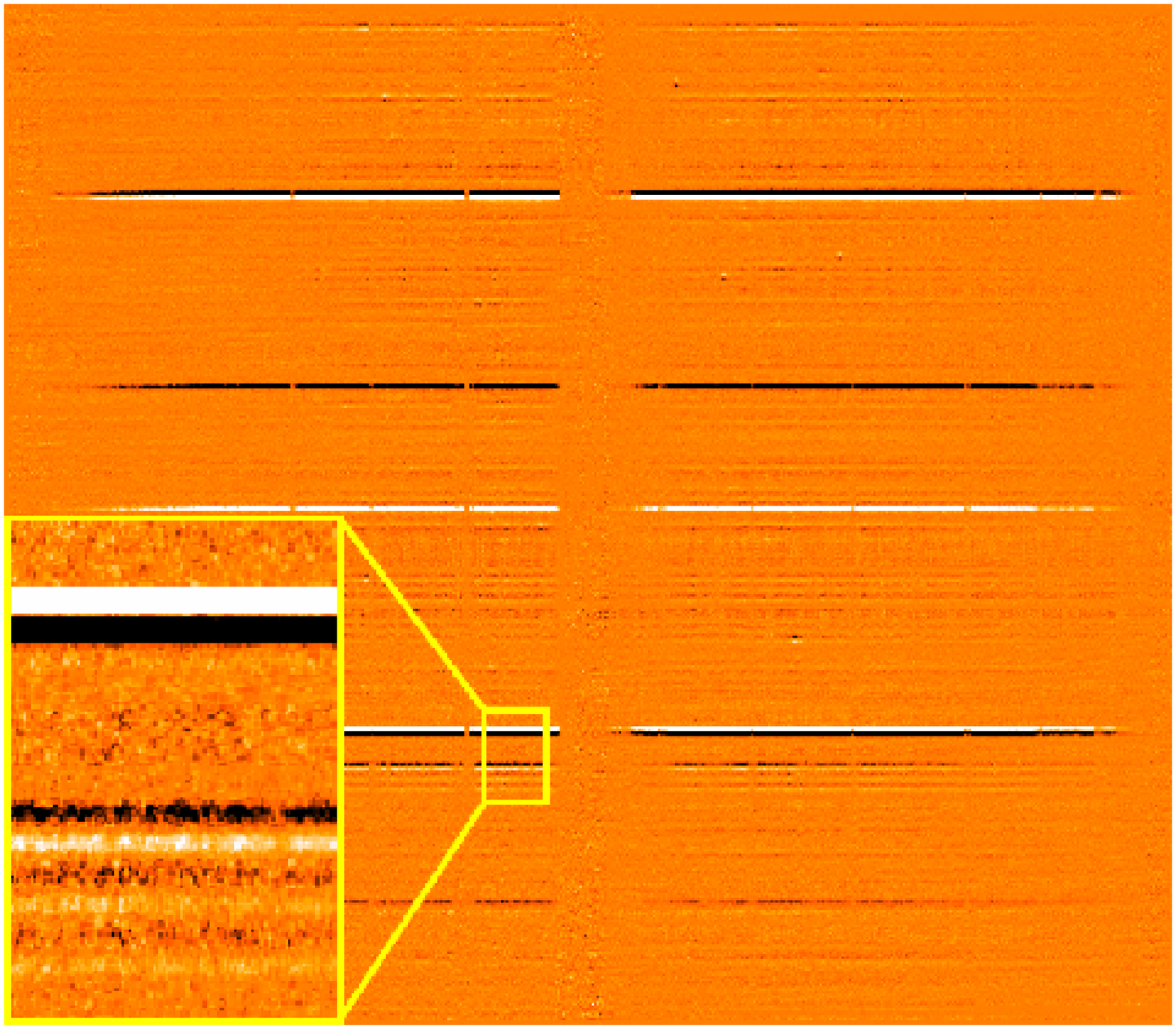}
  \end{center}
  \caption{Combining a pair of target spectra taken in the CBS observation mode. Left panel: Final $A-B$ image in which the positions of the targets observed only at position $B$ are masked. Center panel: Inverted and rearranged negative spectrum that will be combined with the corresponding positive spectrum. Right panel: Combined spectrum. The S/N ratios of the positive spectra are 1.4 times higher than the ratio of the corresponding negative spectra in the combined image. 
}
\label{fig:CBS_combine}
\end{figure*}

Although the correction for spectral distortions was performed to straighten the images of the airglow masks (as in figure \ref{fig:spectral_distortion}), there remain small differences among the throughput patterns of the spectra because of the local shape errors of the mask elements or because of the presence of dust on the mask mirror. These differences are corrected by dividing the averaged image by a ``relative'' dome-flat image in which the common spectral features have been removed by normalizing the count along lines and columns (cf. figure \ref{fig:relative_domeflat}). After correcting the relative differences of the throughput patterns among the spectra, the corresponding spectra from positions $A$ (positive) and $B$ (negative) taken in the CBS observation mode are ready to be combined. The negative spectra in the image are then inverted and rearranged so that they can be combined with the corresponding positive spectra (as shown in figure \ref{fig:CBS_combine}).

\subsection{Object mask}
In the process of fitting and subtract the residual background, some fraction of the object flux can be subtracted. The best way to retain the object flux is to mask the objects during the fit. Faint objects to be masked are therefore selected by a human eye on the combined image and the reduction process is repeated with these masks applied. The mask density should not be too high to make the residual background subtraction work well: the maximum density allowed is roughly 75\% of the image.

\subsection{Square-noise map}
Before the mask edge correction process, the distribution of noise in an image is truncated Poisson distribution caused by the bad pixel rejection process in three subsequent stages. Although the noise level in an image is almost homogeneous at this stage, the noise level map becomes complicated during the throughput correction and CBS combine. To estimate the noise level of each pixel, a frame is defined that consists of the squared noise level measured along columns just before the mask edge correction. Here, the noise level is measured by a 3$\sigma$ clipping algorithm iterated ten times for each column, while some extra contribution proportional to the count is added to the clipped pixels. This square-noise frame is divided by the squared image of the relative dome flat, and reduced to half wherever a pair of spectra is averaged during the CBS combine. Note that the noise level of the bright objects is probably underestimated because the systematic uncertainty (i.e. sub-pixel shifts of the spectra in raw images and tiny variation of the throughput pattern caused by the instrument status such as temperature) is the major contribution for them. 

\section{Flux calibration and check process}
\subsection{Outline}
To calibrate the flux of the scientific targets, at least one bright ($J(AB)=15$--$18$mag) star in each science frames is needed as a spectral reference, because it must have almost the same atmospheric absorption feature as that for the scientific targets in the same field of view. All the spectra are divided by the reference spectrum, and then multiplied by the expected spectrum of the reference star whose flux and spectral type are known or can be determined by the observed values in two different wavelength regions. If the cataloged or estimated spectrum of the reference star is correct, all the observed spectra will then be calibrated accurately. In the next subsection, the method to estimate the reference star spectrum is described.

\subsection{Template stellar spectra}
\begin{figure}
  \begin{center}
    \FigureFile(80mm,80mm){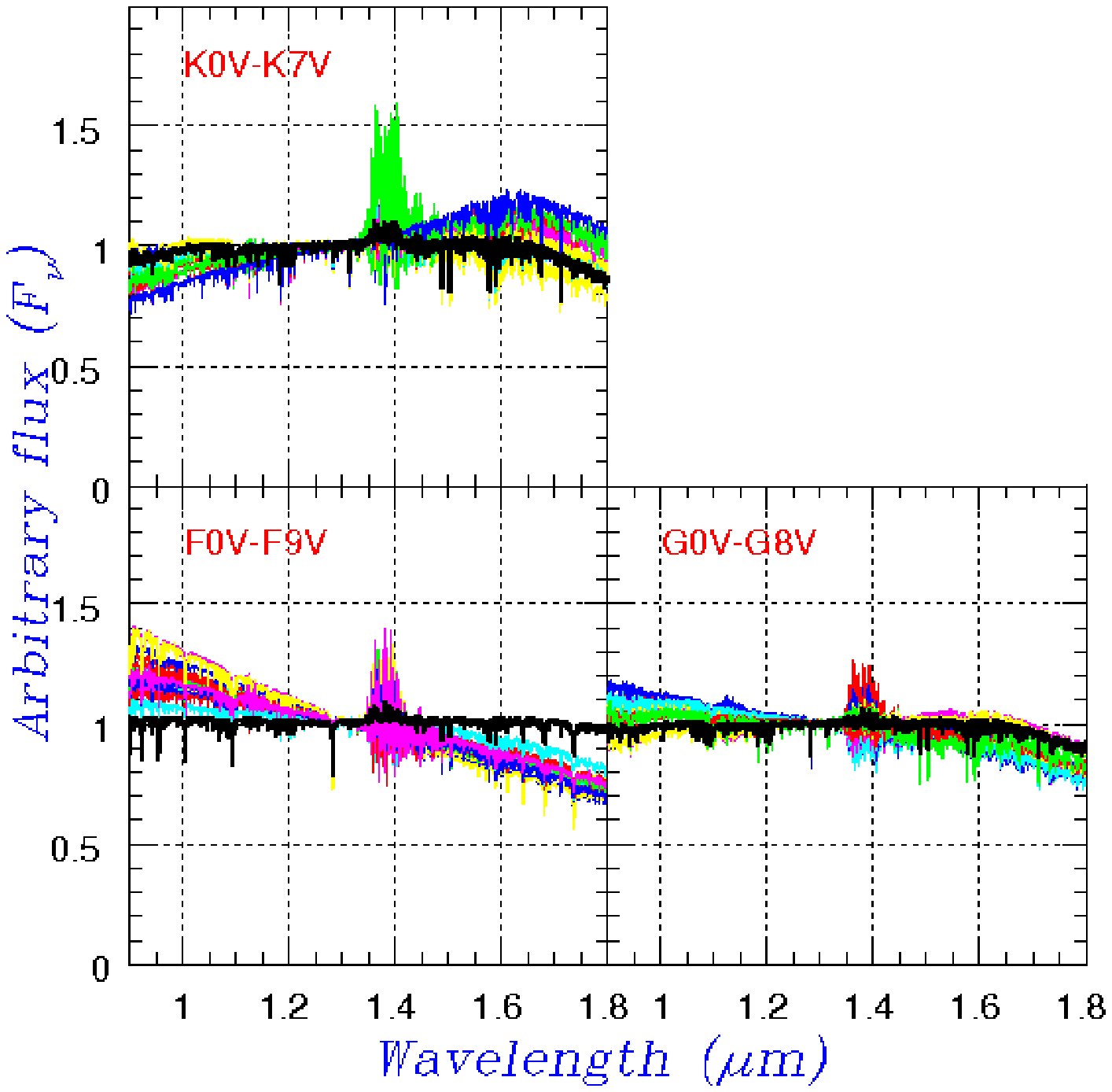}
    \FigureFile(80mm,80mm){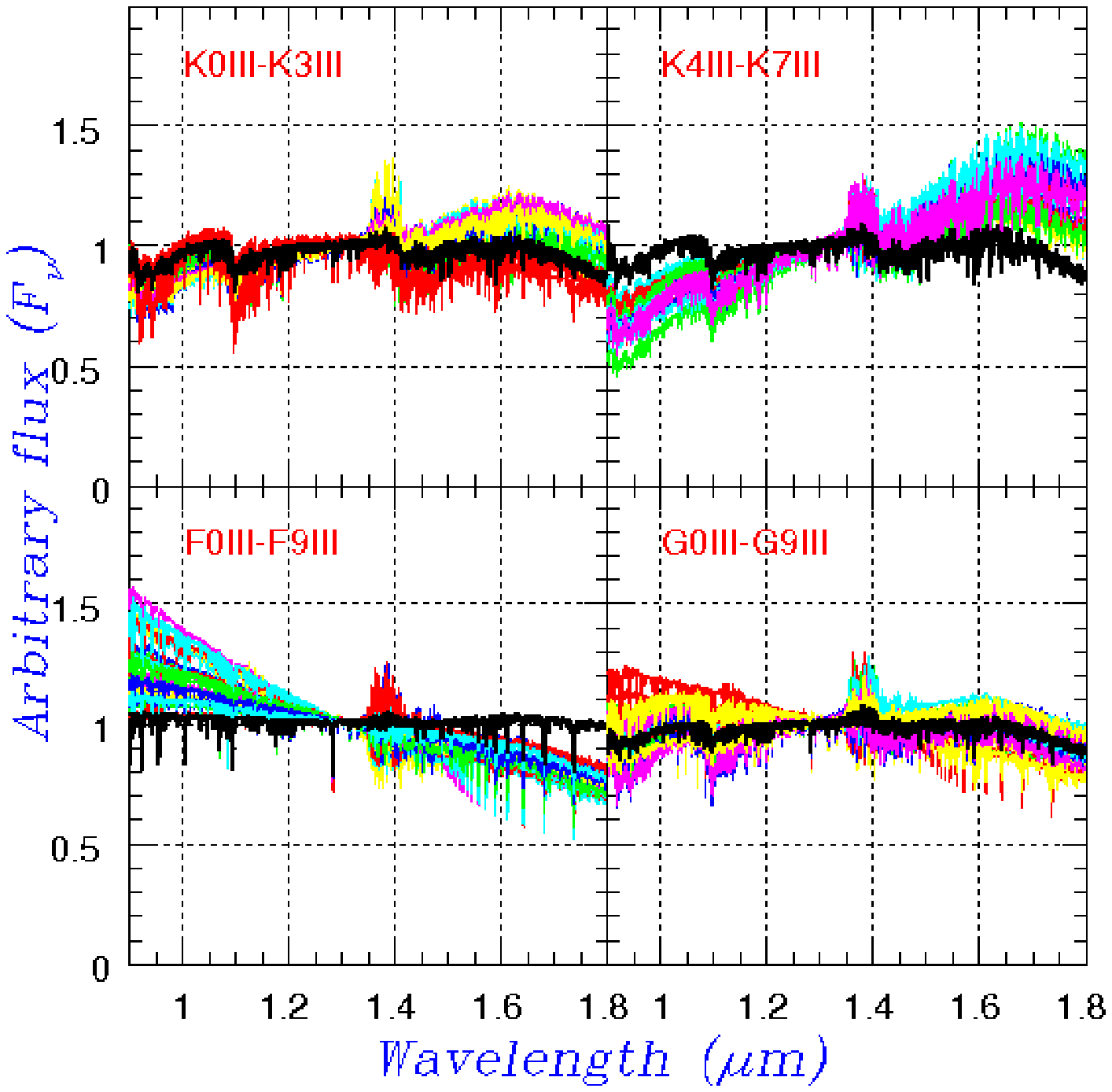}
  \end{center}
  \caption{Stellar spectra in the IRTF Spectral Library. The slope-removed averaged spectrum of each stellar type (thick black lines) is used as the intrinsic absorption template.}\label{fig:irtflib}
\end{figure}

Since the flux and slope of the spectrum of a reference star are determined from the measured counts in an image, one needs the slope-removed template spectra including the intrinsic absorption features of the star. We analyzed 128 stellar spectra in the IRTF Spectral Library\footnote{http://irtfweb.ifa.hawaii.edu/spex/IRTF\_Spectral\_Library/References.html} \citep{ray09} split into seven groups: F0-F9IV/V (19 objects), G0-G8IV/V (14 objects), K0-K7IV/V (10 objects), F0-F9I/II/III (21 objects), G0-G9I/II/III (31 objects), K0-K3I/II/III (17 objects), and K4-K7I/II/III (16 objects). In each group, the spectra were averaged to improve the S/N ratio and divided by the results of the linear fitting, so that slope-removed template spectra of seven different types were prepared (cf. figure \ref{fig:irtflib}). Here, only F, G, and K type stars were used to make the slope-removed template spectra, because 1) their slope-removed spectra become roughly straight in $\lambda$ - $F_{\nu}$ plot, 2) these stars are quite popular and easy to select in the target field, and 3) A and earlier type spectra are neither available in the IRTF Spectral Library, nor in other database with similar qualities.

\begin{figure}
  \begin{center}
    \FigureFile(80mm,80mm){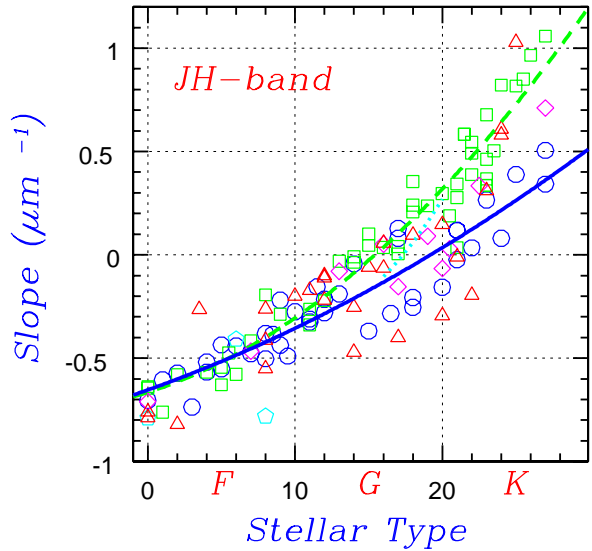}
  \end{center}
  \caption{Correlation between a stellar type and the slope of its spectrum. The stellar types are numbered from 0 (F0) to 29 (K9), and the slope is defined by equation (1). The open symbols are measured from figure \ref{fig:irtflib}: type I (triangles), type II (diamonds), type III (squares), type IV (pentagons), and type V (circles). The distributions of types III and V are fit to second-order polynomials represented by the thick solid and dashed lines, respectively. The thick dotted line indicate the intermediate correlation between types III and V used when the stellar type of a reference star is unknown.}\label{fig:type_slope}
\end{figure}

Next, the correlations between the spectral types and the slopes of the spectra have to be established. Figure \ref{fig:type_slope} shows the distribution of measured value of 128 stellar spectra from figure \ref{fig:irtflib}. The stellar types are numbered from 0 (F0) to 29 (K9) while the slopes are defined by
\begin{equation}
slope = \frac{F_{\nu}(1.55)-F_{\nu}(1.21)}{(1.55-1.21)F_{\nu}(1.31)}
\end{equation}
in $\mu{\rm m}^{-1}$. The correlations between the spectral types and the slopes are determined by second-order polynomial fits to the distributions of stellar types III and V:

\[
slope_{\rm III} = 0.00132type^2+0.0229type-0.666 ,
\]
\begin{equation}
slope_{\rm V} = 0.000456type^2+0.0253type-0.654 .
\end{equation}

As a result, any spectrum from F0 to K9 can be synthesized by multiplying a linear spectrum having a defined slope with the intrinsic absorption spectrum from the interpolation of the nearest two slope-removed template spectra.

\subsection{Calibration process}
The first step in the calibration process is the relative throughput correction of fibers in which the averaged throughput of position $A$ and $B$ is used for the merged spectra in the CBS combine process. The next step is to remap the pixels with an increment of 5 \AA/pixel (1.25 \AA/pixel in the high-resolution mode) based on the correlation between the observed wavelength and the pixels determined in the preparatory reduction process. In this remapping process, the observed wavelengths are multiplied by the count of each pixel in order to convert the value from photon count to $F_{\nu}$. Next, the converted count is divided by the atmospheric transmittance function of the airmass present during the observation, so as to roughly correct for the effects of atmospheric absorption. After this correction, the flux of the reference star is estimated from the converted count at around 1.31$\mu$m, assuming that the total system efficiency under good seeing condition is 2.5\% including losses at the entrance of the fibers. The slope of the spectrum is then measured with a fixed efficiency ratio between 1.21 and 1.55 $\mu$m. (The value of this ratio will be confirmed in the check process, along with the total system efficiency of 2.5\%.) Finally, the observed scientific spectra are divided by the reference spectrum and multiplied by the stellar spectra from the measured flux and slope. Here, the type V and III absorption templates are adopted as the reference spectra, respectively bluer than G5 and redder than K1 (cf. figure \ref{fig:type_slope}). As a consequence, two of the intrinsic absorption templates of FV, GV, K1III, and K4III (in figure \ref{fig:irtflib}) are used to interpolate the absorption of the reference spectrum. If the stellar type of the reference star is known, the corresponding synthesized spectrum is used instead of this predicted spectrum.

\begin{figure*}
  \begin{center}
    \FigureFile(80mm,80mm){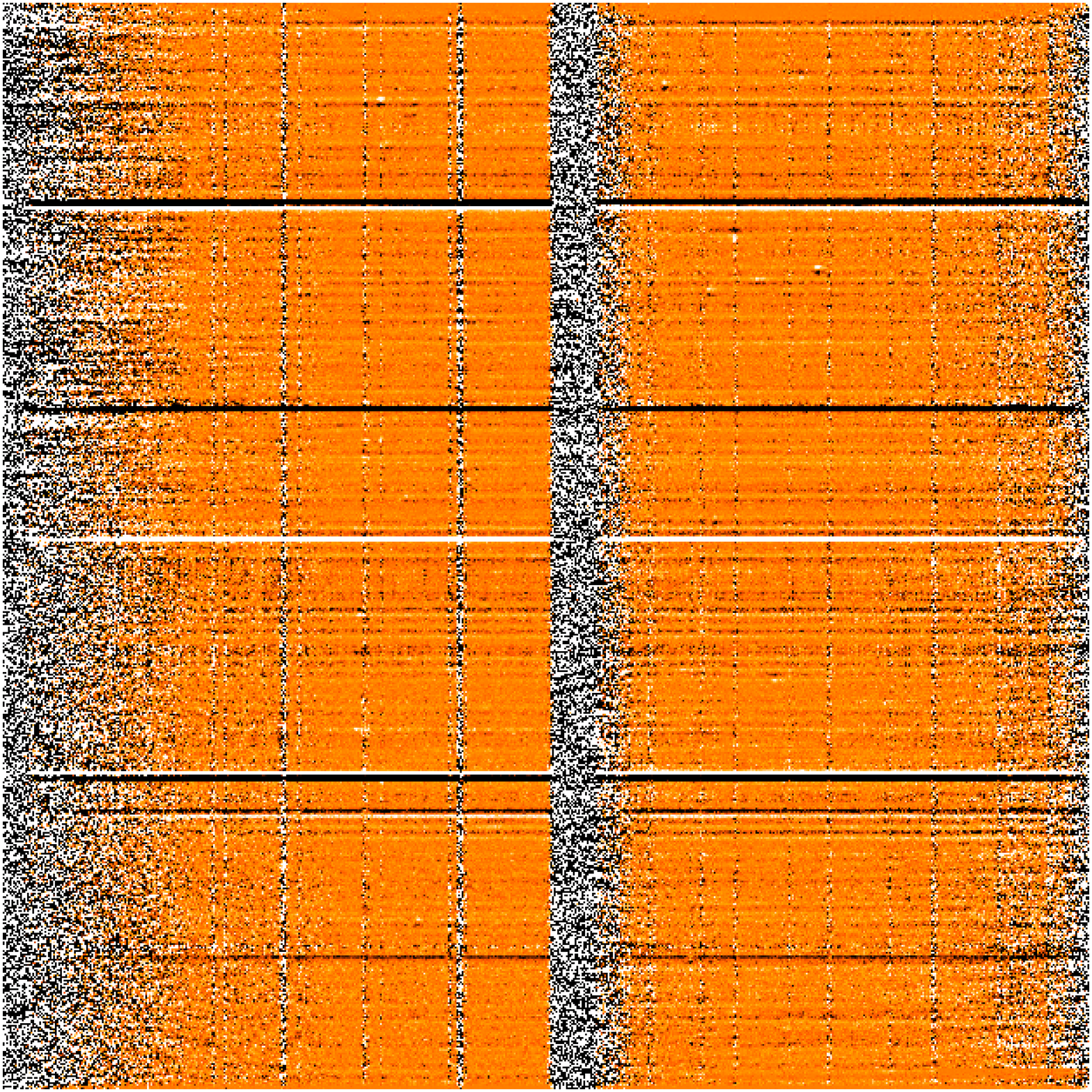}
    \FigureFile(80mm,80mm){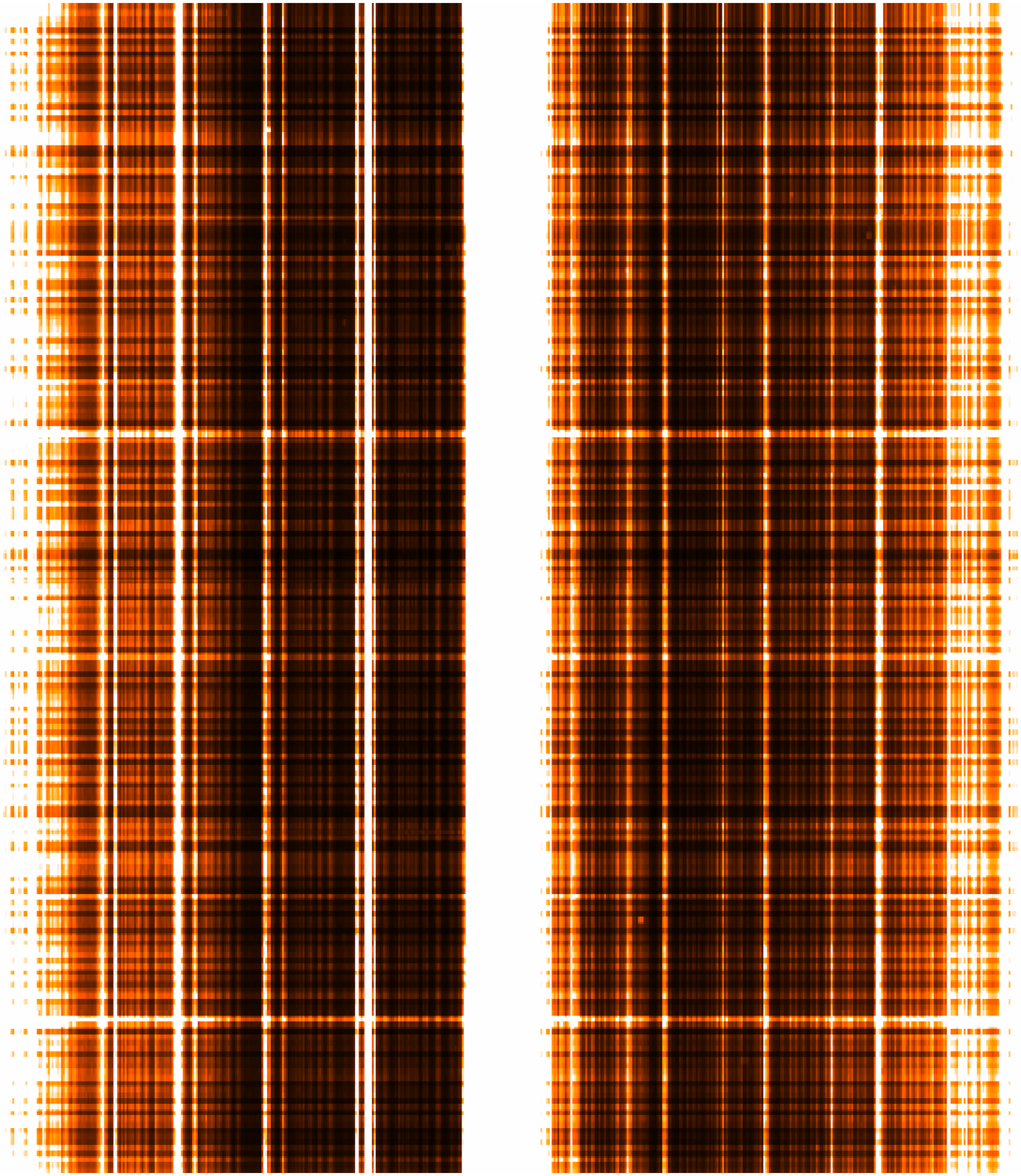}
  \end{center}
  \caption{Calibrated spectral image (at left) and a square-noise frame (at right) with a size of 1800$\times$1800 pixels. The wavelength range is 0.9-1.8 $\mu$m with an increment of 5 \AA/pixel. Each of the 200 spectra has a width of 9 pixels.
}\label{fig:calib_image}
\end{figure*}

\begin{figure}
  \begin{center}
    \FigureFile(80mm,80mm){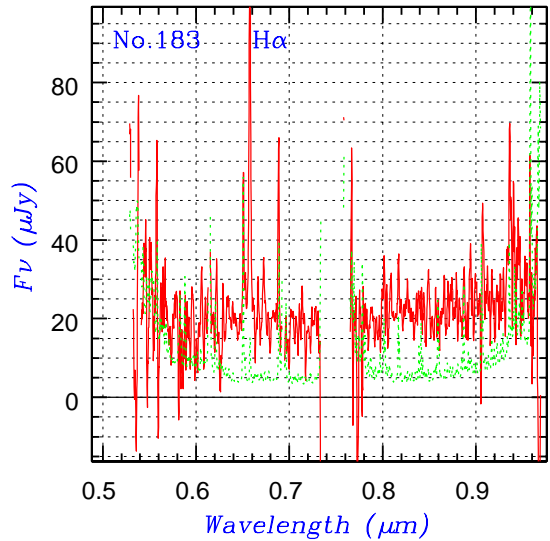}
  \end{center}
  \caption{Example of a processed 1D spectrum. The spectrum and the corresponding 1$\sigma$ noise level are represented by the thick continuous line and the thin dotted line, respectively. The wavelength has been converted to the rest wavelength using the cataloged redshift value ($z=0.844$).}\label{fig:1D_spectrum}
\end{figure}

A resulting wavelength- and flux-calibrated image is shown in figure \ref{fig:calib_image}. The size of this image is 1800$\times$1800 pixels, the wavelength in $\mu$m is $\lambda=0.9+0.0005\times (ColumnNumber-1)\ $, the $n$-th spectrum is located between $y=9\times (LineNumber)-8$ and $y=9\times (LineNumber)$, and the count is in $\mu$Jy. The square-noise frame is converted in a similar way except that all factors are multiplied twice. The 1D spectrum of each object is extracted from this image with a user-defined mask of 9 pixels, together with the square-noise frame. An example of the final 1D spectrum is shown in figure \ref{fig:1D_spectrum}.

\subsection{Check process}
\begin{figure}
  \begin{center}
    \FigureFile(80mm,80mm){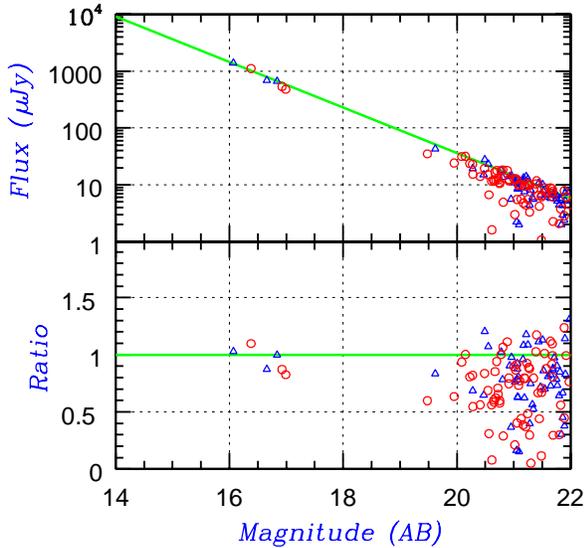}
  \end{center}
  \caption{Comparison of the observed flux with the catalog magnitude. The triangles and circles represent the values in the $J$ and $H$ band, respectively. The thick solid lines indicate the ideal values without loss of flux.}\label{fig:mag_fnu}
\end{figure}

The results are checked by comparing the resulting flux in the $J$ and $H$ bands with the photometric data in the catalog. If all the factors contributing to the system efficiency are normal, the flux should be consistent with the catalog value. The accuracy of the efficiency ratio between 1.21 and 1.55 $\mu$m can also be confirmed by these diagrams for most cases. Figure \ref{fig:mag_fnu} shows an example of such a comparison. However, weather conditions, telescope and instrument focus, extended (non-stellar) morphology of targets, and position accuracy of the catalog may cause the lower observed flux of some targets than that expected from the catalog. Under sufficiently good conditions, the observed flux should match the catalog magnitude as shown in the thick line in figure \ref{fig:mag_fnu} with some downward scatter due to small allocation error of the fibers. Also, if the estimated efficiency ratio between 1.21 and 1.55 is not correct, the points of $J$ and $H$ will have small offset in this figure.

\section{Summary and Conclusions}
The reduction process for FMOS images includes two special processing steps. One is the segmented processing of the spectra to handle a given part of the PSF as a unit, while the other is the automatic modeling of the reference spectrum to calibrate the scientific targets. The segmented processing enable to keep the original 2-dimensional information which has a large effect on bad pixel filtering and detection of faint emission-lines. Most of the processes are carried out automatically, but the object mask preparation and the reference star selection require user judgement. The FIBRE-pac is available from the FMOS instrument page\footnote{http://www.naoj.org/Observing/Instruments/FMOS/} of the Subaru web site, together with the sample dataset presented in this paper. The base reduction platform is IRAF and the reduction scripts and sources are free and open, so that users can check what is happening in each step by sending the commands one at a time.

\bigskip

This work was supported by a Grant-in-Aid for Scientific Research (B) of Japan (22340044) and by a Grant-in-Aid for the Global COE Program "The Next Generation of Physics, Spun from Universality and Emergence" from the Ministry of Education, Culture, Sports, Science, and Technology (MEXT) of Japan.

IRAF is distributed by the National Optical Astronomy Observatories, which are operated by the Association of Universities for Research in Astronomy, Inc., under cooperative agreement with the National Science Foundation.

\end{document}